\documentclass[aps,prl,groupedaddress,showpacs]{revtex4}
\usepackage{amsfonts}
\usepackage{graphicx}

\begin{document}

%\preprint{}

\title{K\"allen-Lehman Representation and the Gluon Propagator}


\author{Marco Frasca}
\email[]{marcofrasca@mclink.it}
%\thanks{}
\affiliation{Via Erasmo Gattamelata, 3 \\ 00176 Roma (Italy)}


\date{\today}

\begin{abstract}
We exploit the K\"allen-Lehman representation of the two-point Green function to
prove that the gluon propagator cannot go to zero in the infrared limit. We
are able to derive also the functional form of it. This means that current 
results on the lattice can be used to derive the scalar glueball spectrum to be
compared both with experiments and different aimed lattice computations.
\end{abstract}

%\keywords{}

\maketitle


Different approaches have been recently devised to extract the behavior of Yang-Mills theory in the infrared limit. These can be mainly listed in two principal categories: Numerical and phenomenological. Numerical approaches manage Yang-Mills quantum field theory on the lattice or just try to solve the Dyson-Schwinger equations. Phenomenological approaches try to extract from experimental data the behavior of the propagator or the running coupling in the same limit. 

Recent results from both approaches are outlining a clear scenario. Lattice results are showing that the gluon propagator tends to a non null value in the infrared, the ghost propagator is the free particle propagator and the
running coupling goes to zero in the same limit \cite{ilg1,ste1,cuc1}. Solution of the Dyson-Schwinger equations is showing a similar scenario in 4d \cite{nat1} while on a torus the running coupling is seen to go to zero while for the propagator the results are not so clear \cite{fis1}. On the phenomenological side there is evidence that the gluon propagator should not tend to zero \cite{nat2} and that the running coupling goes to zero \cite{pro}.

The extraction of a mass gap from experimental results is somewhat ambiguous while for lattice computations is rather straightforward. Indeed, on the lattice for the spectrum of glueballs the results are widely known \cite{tep1,morn}. The problem with lattice computations in this case is how the coarsening affects the determination of the true ground state of the theory. It is a well known fact that the f$_0$(600) meson is not seen in any lattice computation both for Yang-Mills and full QCD. Indeed, the nature of this resonance is a strongly debated matter in the literature and is strongly linked to the fact if this is just a quark composite meson or not. In this latter case it should be a glueball and the true ground state of the Yang-Mills theory being the lightest meson without quark content. In a full QCD computation, mixing at higher energy with quark states should be anyhow expected. An indication that this is the correct scenario has been recently shown using experimental results \cite{pel1,och}. Recently, authors also improved the mass determination of this meson \cite{pel2}. They obtain $484 \pm 17$MeV and this result should be compared with the Yang-Mills integration constant generally used in lattice computations 440 MeV, derived from experimental results, that has a ratio with it of 1.1. We can conclude that, although the exact value of the ground state energy is to be determined, it is proven without doubt both by lattice computations and experimental results that a quantum Yang-Mills theory has a mass gap in the low energy limit (infrared limit). 

So, we have a complete scenario for the behavior of a  quantum Yang-Mills theory in the infrared. We can summarize all these facts as follows. Firstly we introduce the gluon propagator as
\begin{equation}
    D(p) = \frac{Z(p)}{p^2},
\end{equation}
the ghost propagator as
\begin{equation}
    D^G(p) = -\frac{G(p)}{p^2},
\end{equation}
and the running coupling as
\begin{equation}
    \alpha(p) = G^2(p)Z(p).
\end{equation}
Then we have that
\begin{equation}
    \lim_{p\rightarrow 0} D(p) = constant
\end{equation}
and
\begin{equation}
    \lim_{p\rightarrow 0} D^G(p) = \infty
\end{equation}
and
\begin{equation}
    \lim_{p\rightarrow 0} \alpha(p) = 0
\end{equation}
while the theory shows a mass gap, that is, the gluon propagator should have a pole for some value of the mass $m_0$
such that $p^2=-m^2_0$.

The K\"allen-Lehman representation does hold also without the axiom of positivity. This can be seen by a formulation of quantum field theories without positivity axiom, particularly needed for gauge theory and confinement, that has been obtained by Strocchi e Jakobczyk \cite{stro1,stro2}.

From the K\"allen-Lehman representation \cite{stro2,wei1} one has
\begin{equation}
    D(p) = \frac{Z_0}{p^2+m_0^2-i\epsilon}+\int_{m_1^2}^\infty d\mu^2 \sigma(\mu^2)\frac{1}{p^2+\mu^2-i\epsilon}
\end{equation}
with $m_0<m_1$ and $Z_0$ a renormalization constant. It is straightforward to obtain the well-known result that
should be satisfied by any well-behaving theory that
\begin{equation}
    D(p)\stackrel{p\rightarrow\infty}{=}\frac{1}{p^2}\left(Z_0+\int_{m_1^2}^\infty d\mu^2 \sigma(\mu^2)\right)
\end{equation}
giving immediately
\begin{equation}
\label{eq:Z}
    Z_0+\int_{m_1^2}^\infty d\mu^2 \sigma(\mu^2)=1.
\end{equation}
We know anyhow that the theory in the ultraviolet limit needs some renormalization procedure to recover this
latter equality being the constant $Z_0$ infinite in this case.

From the K\"allen-Lehman representation we also know that $\sigma(0)=0$ unless the theory has some massless
excitation that we have already excluded from the start. So, one has
\begin{equation}
    D(p)\stackrel{p\rightarrow 0}{=}\frac{Z_0}{m_0^2}+\int_{m_1^2}^\infty d\mu^2 \frac{\sigma(\mu^2)}{\mu^2}
\end{equation}
but there is no pole at $\mu=0$ in the integration range and this finally means
\begin{equation}
    D(p)\stackrel{p\rightarrow 0}{=} constant
\end{equation}
where the constant is greater than $\frac{Z_0}{m_0^2}$ different from zero. 

This result is very important in view of the fact that it permits to connect two different kinds of analyses generally carried out with the theory on the lattice. Indeed, we can use the above result to see that, after a Wick rotation,
one has
\begin{equation}
    D(0,t) \approx A_0 e^{-m_0 t}
\end{equation}
in the limit $t\rightarrow\infty$ and so the mass gap of the theory can be derived from the gluon propagator computed on the lattice. This result should be compared with the one obtained in \cite{tep1,morn}. We notice that the spectrum of the theory is generally obtained on the lattice assuming that
\begin{equation}
   D(0,t) = \sum_n A_n e^{-m_n t}.
\end{equation}
Indeed, we note that the above procedure can be iterated by the further assumption that the theory has a discrete spectrum. This means that
\begin{equation}
   D(p)=\sum_n\frac{Z_n}{p^2+m_n^2-i\epsilon}
\end{equation}
with $\sigma(\mu^2)=\sum_n Z_n\delta(\mu^2-m^2_n)$ having the form of a typical density of states proper to condensed matter physics for a discrete spectrum.  If one assumes the positivity condition for the spectral representation then all $Z_n$s should be positive. If this is not the case one should maintain condition (\ref{eq:Z}) by assuming $Z_n$ having alternating signs. This is due to the fact that one wants to recover the proper ultraviolet limit. On this basis, the full scalar glueball spectrum is obtainable in principle from the lattice computation of the gluon propagator. This form of the propagator deduced from the K\"allen-Lehman representation of the exact propagator is consistent with the one obtained in \cite{fra1,fra2,fra3}.

We now try to compare this form of the propagator given by the K\"allen-Lehman representation with the scenario depicted above. In order to do this we do the ansatz that holds in the infrared limit \cite{fis2}
\begin{equation}
   Z(p) = c_D(p^2)^{\kappa_D}
\end{equation}
and
\begin{equation}
   G(p) = c_G(p^2)^{\kappa_G}
\end{equation}
and we determine the exponents $\kappa_D$ and $\kappa_G$. For the gluon propagator, for the theorem we proved above, one has immediately that $\kappa_D=1$ and $c_D\ne 0$. In order to evaluate the exponent for the ghost propagator we turn our attention to the running coupling. In the infrared limit we should have
\begin{equation}
   \alpha(p) = c_Dc_G^2 (p^2)^{2\kappa_G+1}
\end{equation}
from which we derive that it is enough to have $2\kappa_G+1 > 0$ to be consistent with lattice and experimental results. If the ghost behaves as a free particle one will have $\kappa_G=0$ and this is the result that is going to emerge into lattice computations. This result is in agreement with the recent work \cite{che} further supporting this scenario depicted in \cite{fra1,fra2,fra3}.

\begin{figure}[tbp]
\begin{center}
\includegraphics[angle=0,width=240pt]{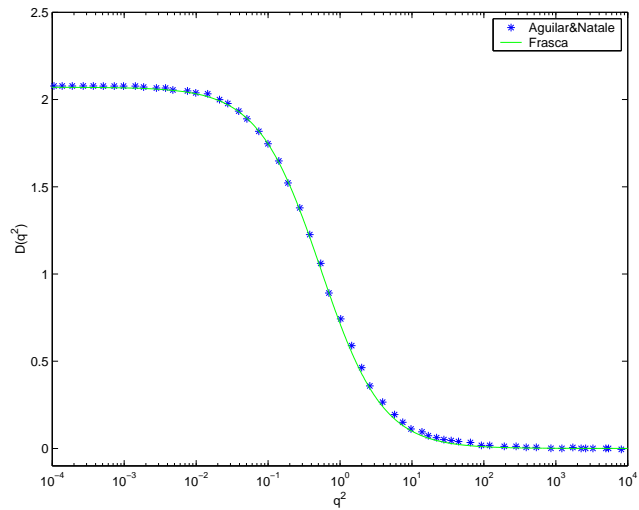}
\caption{\label{fig:fig1} Aguilar and Natale numerical solution of Dyson-Schwinger equations for Yang-Mills theory \cite{nat1} compared to our propagator in agreement with K\"allen-Lehman representation\cite{fra1,fra2,fra3}.}
\end{center}
\end{figure}

% Numerical comparison
We would like to verify that our idea of a K\"allen-Lehman representation is indeed in agreement with numerical results.
Aguilar and Natale \cite{nat1} numerically solved Dyson-Schwinger equations obtaining a scenario in agreement with the one presented here and recent lattice computations. As this represents a continuum solution for Yang-Mills theory we need to get a perfect agreement by our form of the propagator that agrees with K\"allen-Lehman representation \cite{fra1,fra2,fra3}
\begin{equation}
    D(p)=\sum_{n=0}^\infty\frac{B_n}{p^2+m_n^2-i\epsilon}
\end{equation}
being
\begin{equation}
    B_n=(2n+1)\frac{\pi^2}{K^2(i)}\frac{(-1)^{n+1}e^{-(n+\frac{1}{2})\pi}}{1+e^{-(2n+1)\pi}},
\end{equation}
and
\begin{equation}
    m_n = (2n+1)\frac{\pi}{2K(i)}\sqrt{\sigma},
\end{equation}
and the numerical solution. Here $\sigma$ is the string tension to be fixed experimentally. The result of the comparison is shown in fig.\ref{fig:fig1} and is exceptionally good for a gluon mass of 738 MeV.
This means that a K\"allen-Lehman type of gluon propagator fits very well the numerical solution of the Yang-Mills equations while describes correctly the scenario that is now emerging from lattice computations.

One may ask what should be the next step after a proper determination of the propagator of the Yang-Mills theory in the infrared, being this consistent with the K\"allen-Lehman representation. Indeed, one can do scattering theory by using Lehman-Symanzyk-Zimmermann reduction formula making in this way a Yang-Mills theory manageable also in this limit of a strong coupling. As for the ultraviolet case, we should be granted that the renormalization constants that are computed in this limit are finite. Presently, lattice computations and theoretical analysis point toward this direction making possible to get finally a full theory to carry out further computations to be compared with experiments.

We have proven that the gluon propagator is not zero in the limit of zero momentum using the K\"allen-Lehman representation. We have shown in this way how lattice computations, done to compute the propagator, can be used to obtain the spectrum of the theory assuming this spectral representation to hold also if the spectral density is not always positive. The spectrum is generally obtained by lattice computations but with other techniques. The results given by the two approaches can be compared in this way. This very simple computation to be accomplished should give really striking results producing a serious consistency check.

\begin{acknowledgments}
I would like to thank Valter Moretti for very helpful comments about K\"allen-Lehman representation.
\end{acknowledgments}

\end{document}